\documentclass[twocolumn,prx,superscriptaddress]{revtex4-1}
\newcommand{\pagenumbaa}{1}
\usepackage{graphicx}
\usepackage{amsmath}
\usepackage{siunitx}
\usepackage{hyperref}
\usepackage{color}

\begin{document}

\title{Stretchable persistent spin helices in GaAs quantum wells}

\author{Florian Dettwiler}
\affiliation
{Department of Physics, University of Basel, CH-4056, Basel, Switzerland}

\author{Jiyong Fu}
\thanks{Permanent address: Department of Physics, Qufu Normal University,
  Qufu, Shandong, 273165, China}
\affiliation
{Instituto de F\'isica de S\~ao Carlos, Universidade de S\~ao Paulo,
13560-970 S\~ao Carlos, SP, Brazil}

\author{Shawn Mack}
\thanks{Current address: Naval Research Laboratory, Washington, DC 20375, USA}
\affiliation
{California NanoSystems Institute, University of California, Santa Barbara,
California 93106, USA}

\author{Pirmin J. Weigele}
\affiliation
{Department of Physics, University of Basel, CH-4056, Basel, Switzerland}

\author{J. Carlos Egues}
\affiliation
{Instituto de F\'isica de S\~ao Carlos, Universidade de S\~ao Paulo,
13560-970 S\~ao Carlos, SP, Brazil}

\author{David D. Awschalom}
\affiliation
{California NanoSystems Institute, University of California, Santa Barbara,
California 93106, USA}
\affiliation
{Institute for Molecular Engineering, University of Chicago, Chicago, IL  60637 USA}

\author{Dominik M. Zumb\"uhl}
\affiliation
{Department of Physics, University of Basel, CH-4056, Basel, Switzerland}
\date{\today}
\begin{abstract}
The Rashba and Dresselhaus spin-orbit (SO) interactions in 2D electron gases act as effective magnetic fields with momentum-dependent directions, which cause spin decay as the spins undergo arbitrary precessions about these randomly-oriented SO fields due to momentum scattering. Theoretically  and experimentally, it has been established that by fine-tuning the Rashba $\alpha$ and Dresselhaus $\beta$ couplings to equal {\it fixed} strengths $\alpha=\beta$, the total SO field becomes unidirectional thus rendering the electron spins immune to dephasing due to momentum scattering. A robust persistent spin helix (PSH), i.e., a helical spin-density wave excitation with constant pitch $P=2\pi/Q$, $Q=4m\alpha/\hbar^2$, has already been experimentally realized at this singular point $\alpha=\beta$.  Here we employ the suppression of weak antilocalization as a sensitive detector for matched SO fields together with a technique that allows for independent electrical control over the SO couplings via top gate voltage $V_T$ and back gate voltage $V_B$, to extract all SO couplings as functions of $V_T$ and $V_B$ when combined with detailed numerical simulations. We demonstrate for the first time the gate control of $\beta$ and the {\it continuous locking} of the SO fields at $\alpha=\beta$, i.e., we are able to vary both $\alpha$ and $\beta$ controllably and continuously with $V_T$ and $V_B$, while keeping them locked at equal strengths.  This makes possible a new concept: ``stretchable PSHs'', i.e., helical spin patterns with continuously variable pitches $P$ over a wide parameter range. This further protects spins from decay when electrically controlling the spin precession.
We also quantify the detrimental effect of the cubic Dresselhaus term, which breaks the unidirectionality of the total SO field and causes spin decay at higher electron densities. The extracted spin-diffusion lengths and decay times as a function of  $\alpha/\beta$ show a significant enhancement near $\alpha/\beta=1$. Since within the continuous-locking regime quantum transport is diffusive (2D) for charge while ballistic (1D) for spin and thus amenable to coherent spin control, stretchable PSHs could provide the platform for the much heralded
long-distance communication $\sim 8 - 25$ $\mu$m between solid-state spin qubits, where the spin diffusion length for $\alpha \neq\beta$ is an order of magnitude smaller. 
%
\end{abstract}
\maketitle

\setcounter{page}{\pagenumbaa}
\thispagestyle{plain}


\bfseries






\normalfont


The inextricable coupling between the electron spatial and spin degrees of freedom -- the spin-orbit (SO) interaction --   underlies many fundamental phenomena such as the spin Hall effects -- quantum and anomalous~\cite{sinova:2015} -- and plays a crucial role in newly discovered quantum materials hosting Majorana~\cite{mourik:2012} and Weyl fermions~\cite{wan:2011}.
In nanostructures the SO coupling strength can be varied via gate electrodes~\cite{engels:1997,nitta:1997}.  As recently demonstrated~\cite{chuang:2015}, this enables controlled spin modulation \cite{datta:1990} of charge currents in non-magnetic (quasi-ballistic) spin transistors.
%

\begin{figure}[t]
\includegraphics[width=8.7cm]{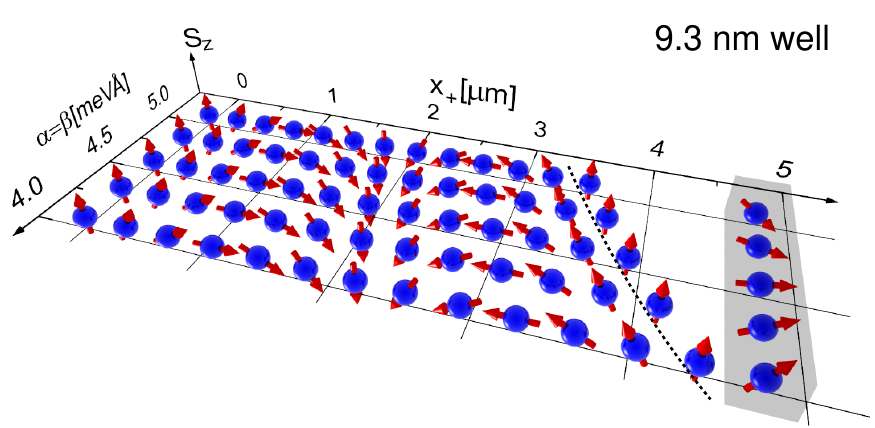}
\vspace{-6mm}
\caption{{\bf Stretchable PSHs}. Illustration of spin helices at different values of $\alpha=\beta$ accessible in the measurements. The position $x_+$ for one $2\pi$ rotation is changing for the gate-locked regime $\alpha=\beta$ (indicated by the the dashed curve). The grey box highlights how the spin rotation can be controlled (in-situ) at fixed position $\sim$5\,$\rm{\mu m}$ over the same range of $\alpha=\beta$.  The $\hat{x}_+ || [110]$ and $\hat{x}_- || [\bar{1}10]$ axes define the 2D plane.}
\label{fig:0}\vspace{-6mm}
\end{figure}

The SO coupling in a GaAs quantum well has two dominant contributions: the Rashba~\cite{rashba:1984} and the Dresselhaus~\cite{dresselhaus:1955} effects, arising from the breaking of the  structural and crystal inversion symmetries, respectively.  When the Rashba $\alpha$ and Dresselhaus $\beta$ SO couplings match at $\alpha=\beta$~\cite{schliemann:2003,bernevig:2006}, the direction of the combined Rashba-Dresselhaus field becomes momentum independent thus suppressing D'yakonov-Perel and Elliott-Yafet spin-flip processes due to non-magnetic impurities, provided that the cubic Dresselhaus term be small.  The significantly enhanced spin lifetime at $\alpha=\beta$ enables non-ballistic spin transistors and persistent spin helices~\cite{schliemann:2003,bernevig:2006}. However, despite substantial efforts, so far this symmetry point has only been achieved at isolated points  with  finely-tuned system parameters~\cite{koralek:2009,walser:2012_1,kohda:2012}, which is too difficult to be reliably attained {\it on demand} as required for a useful technology.

{\it Stretchable Persistent Spin Helices.---} Here we overcome this outstanding obstacle by (i) using a technique that allows independent control of the SO couplings via  a top gate voltage $V_T$ and a back gate voltage $V_B $, which control the electron density $n$ and the electric fields in the well, while (ii) simultaneously measuring the suppression of weak-antilocalization (WAL) in an external magnetic field as a sensitive probe for the matched SO couplings. We demonstrate a robust {\it continuous locking} of the Rashba and Dresselhaus couplings at $\alpha(V_T,V_B)=\beta(V_T,V_B)$ over a wide range of densities $n$, i.e., a ``symmetry line'' (not a point) in the ($V_T,V_B$) plane. More specifically, for a $9.3$ nm wide GaAs well we can vary the SO couplings continuously and controllably  from  $\alpha=\beta=5\, \mathrm{ meV \AA}$ to $4\,\mathrm{ meV \AA}$. This enables ``stretchable spin helices", see Fig.~\ref{fig:0}, with spin density $s_{x_+}\sim \sin(Qx_+)$, $s_{x_-}=0$, and $s_z\sim \cos(Qx_+)$ and {\it variable} pitches $P=2\pi/Q$, $Q=4m\alpha/\hbar^2$,  that can coherently couple spin qubits over unprecedented long distances.


{\it Long-distance spin communication.---} Within the range of the continuously matched-locked SO couplings $\alpha=\beta$, quantum transport in the well is diffusive for charge (2D) while essentially ballistic (1D) for spins (see SOM Sec.~V).  The cubic Dresselhaus term is small in this range as we quantify later on and leads to spin decay with spin-diffusion lengths $\lambda_{\rm eff} \sim 8-25 \mu$m over which spin dephases by 1 radian. The full electrical control of the SO couplings demonstrated in our 9.3 nm wide quantum well enables stretchable PSHs with pitches $P$ stretching from $3.5\,\mathrm{\mu m}$ to $4.5\,\mathrm{\mu m}$, see Fig.~\ref{fig:0}. These stretchy waves can be excited upon injection of spin polarization, see e.g., Refs.~\cite{koralek:2009,walser:2012_1}. Figure~\ref{fig:0} illustrates how spin information can be conveyed between spins via a stretchable PSH. Within the shortest spin-diffusion length $\lambda_{\rm eff}\sim8\,\mathrm{\mu m}$ for our 9.3 nm well, controlled spin rotations $\theta=Qx_+=2\pi x_+/P$ can be performed under spin protection on any spin sitting at a position $x$ along the stretchable PSH by varying $P$ in the range above. For example, a spin at $x\sim4.5\, \mathrm{\mu m}$ can be rotated by $\Delta \theta\sim \pi/2$ as $P$ varies in the range above, see gray box shading in Fig.~\ref{fig:0}. Other spin communication modes can be envisaged with this setup. Note that this type of spin control, manipulation and spin transfer is not possible for a GaAs  helix  with $\alpha \ll ({\rm or} \gg) \beta$ as $\lambda_{\rm eff} \lesssim 1$ $\mu$m in this case. Stretchable helices could provide a platform for unprecedented long-distance spin communication between spin qubits defined in GaAs 2D gases.



%
{\it Additional results.} WAL was also used to identify other regimes such as the Dresselhaus regime ($\alpha=0$, Fig.~4a) in a more symmetrically doped sample. Combined with numerical simulations, we extracted the SO couplings $\alpha$ and $\beta$, the bulk Dresselhaus parameter $\gamma$, the spin-diffusion lengths and spin-relaxation times over a wide range of system parameters.  We also quantified the detrimental effects of the third harmonic of the cubic Dresselhaus term Fig.~\ref{fig:4}, which limits spin protection at higher densities. Interestingly, our spin diffusion lengths and spin-relaxation times are significantly enhanced within the locked $\alpha=\beta$ range thus attesting that our proposed setup offers a promising route for spin protection and manipulation.

In what follows we first explain the essential density dependence of the Dresselhaus coupling $\beta$ that enables the continuous locking of the SO fields, how it also leads to spin decay at higher densities, and then the relevant WL/WAL detection scheme, measurements and simulations. A full account of our approach, including additional data and details of the model and simulations, is presented in the Appendix and the SOM.


{\it Linear $\&$ cubic Dresselhaus terms in 2D.} Due to the well
confinement along the $z$ direction (growth), the cubic-in-momentum
bulk (3D) Dresselhaus SO interaction gives rise to, after the projection into the lowest
quantum well subband eigenstates, distinct terms that are linear and cubic in
$\mathbf{k}$, the 2D electron wave vector. The linear-in-k term has a coefficient
$\beta_1=\gamma\langle k_z^2\rangle$ and is practically independent of the density in the
parameter range of interest here Fig.~\ref{fig:2}(b),(d), as we discuss below.
The cubic-in-$k$ term, on the other hand, is density dependent and has yet two
components with distinct angular symmetries: (i) the first-harmonic contribution
proportional to $\sin\phi$ and $\cos\phi$ and (ii) the third-harmonic contribution proportional
to $\sin3\phi$ and $\cos3\phi$; here $\phi$ is the polar angle in 2D  between
$\mathbf{k}$ and the [100] direction (see SOM). Interestingly, the first-harmonic
contribution with coefficient $\beta_3$ has the same angular symmetry as both the linear-in-$k$ Dresselhaus
$\beta_1$ term (see Refs. \cite{iordanskii:1994, pikus:1995} and SOM), and the Rashba $\alpha$ term.

To a very good approximation the coefficient $\beta_3\simeq \gamma k_F^2/4$,  where the Fermi vector $k_F\simeq \sqrt{2\pi n}$ and $n$ is the carrier density of the 2D gas. This neglects the tiny angular
anisotropy in the Fermi wave vector due to the competition between the Rashba and Dresselhaus
effects (specially in GaAs wells). Note that by approximating $\beta_3\simeq \gamma \pi n/2$ both the
first-harmonic and the third-harmonic parts of the cubic-in-$k$ Dresselhaus term become actually linear
in $k$ [see SOM, Eqs.~(S20)-(S21)] and, more importantly, become density dependent. We can now group the linear-in-$k$ Dresselhaus term $\beta_1$ together with the first-harmonic contribution $\beta_3$ into a single linear Dresselhaus term by defining $\beta=\beta_1 - \beta_3$ [details are given in the SOM, Eq.~(S15)]. As described below, it is this density-dependent coefficient $\beta$ that can be tuned with a gate voltage to match the Rashba $\alpha$ coupling continuosly, see Fig.~\ref{fig:2}(b). This matching leads to a $k$-independent spinor (or, equivalently, to a $k$-independent effective SO field), whose direction is immune to momentum scattering.

{\it Spin decay at higher densities.} The strength of the third-harmonic contribution of the Dresselhaus term is also described by the coefficient $\beta_3$. This term, however, is detrimental to spin protection as it breaks the angular symmetry of the other linear SO terms and makes the spinor $k$-dependent and susceptible to in-plane momentum scattering, even for matched couplings $\alpha=\beta$. As we discuss later on (Fig. 4), the detrimental effect of the third-harmonic contribution does not prevent our attaining the continuous locking over a relevant wide range of electron densities.


\begin{figure}[t]
\includegraphics[width=8.7cm]{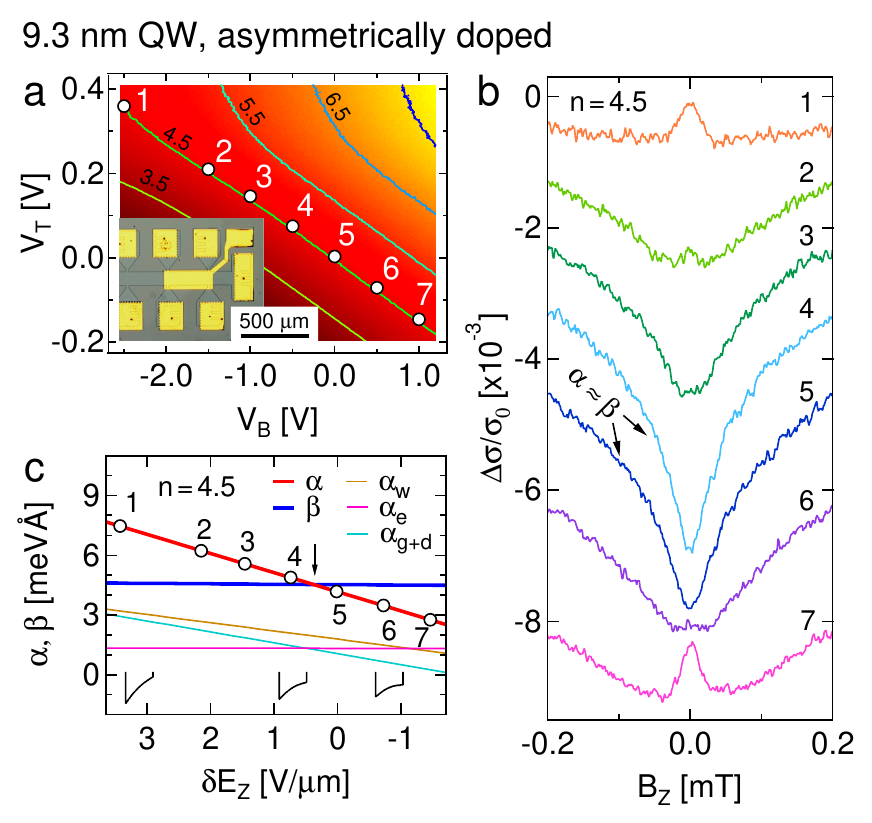}
\vspace{-6mm}
\caption{Weak localization (WL) as an $\alpha=\beta$ detector; gate-control of Rashba $\alpha$ at constant density.  {\bf (a)} Measured charge density $n$ (color) versus top gate voltage $V_T$ and back gate voltage $V_B$ ($9.3\,$nm well). Contours of constant density $3.5 - 7.5\cdot 10^{11}\,\mathrm{cm^{-2}}$ are shown. Inset: optical micrograph of typical Hall bar, with contacts (yellow), gate (center) and mesa (black lines). {\bf (b)} Normalized longitudinal conductivity $\Delta \sigma / \sigma_0=(\sigma(B_Z)-\sigma(0))/\sigma(0)$ versus $B_Z \bot$ 2D plane. Curves for gate configurations $1-7$ along constant $n=4.5\cdot 10^{11}\,\mathrm{cm^{-2}}$ are shown (offset vertically), also labeled in panels {\bf (a)} and {\bf (c)}. {\bf (c)} Simulated Rashba $\alpha$ and Dresselhaus $\beta$ coefficients (see text) against gate-induced field change $\delta E_Z$, shown for constant $n=4.5\cdot10^{11}\,\mathrm{cm^{-2}}$. The $\delta E_Z$ axis -- decreasing from left to right -- corresponds exactly to the $V_B$ abscissa of panel {\bf (a)} for a covarying $V_T$ such that $n=4.5\cdot10^{11}\,\mathrm{cm^{-2}}$ constant. Sketches of the well potential at 1, 4 and 6 illustrate the change of $\alpha$ with $\delta E_Z$. Note that $\alpha(\delta E_Z=0)\neq 0$ since the external E-field (see SOM) is not zero at $\delta E_Z=0$.}
\label{fig:1}\vspace{-6mm}
\end{figure}

{\it Gate-tunable range of the Dresselhaus coupling $\beta$.} For the narrow quantum wells used here, $\beta_1$ is essentially gate-independent since the wave function spreads over the full width of the well. This also implies $\langle k_z^2\rangle\ll(\pi/W)^2$ (the infinite well limit), see Fig.\,\ref{fig:2}d, due to wave function penetration into the finite barriers. Thus, a change of density by a factor of $\sim 2.5$ changes $\beta_3/\beta_1= \pi n/\langle 2 k_z^2\rangle$ by the same factor, resulting in a gate-tunable range of $0.08~\lesssim~\beta_3/\beta_1\lesssim~0.2$. In addition, quantum wells of width $W=8, 9.3, 11$ and $13\,$nm were used\cite{luo:1990,koralek:2009}, resulting in a change of $\beta_1$ by roughly a factor of $2$.

{\it Controlling the Rashba coupling $\alpha$.} The Rashba coefficient~\cite{rashba:1984} $\alpha$ can be tuned with the wafer and doping profile~\cite{koralek:2009} as well as \emph{in-situ} using gate voltages\cite{engels:1997,nitta:1997} at constant density and thus independent of the Dresselhaus term. A change of top gate voltage $V_T$ can be compensated by an appropriate, opposing change of back gate voltage $V_B$ (see Fig.\,\ref{fig:1}a) to keep $n$ fixed~\cite{papadakis:1999,grundler:2000} while changing the gate-induced electric field $\delta E_Z$ in the quantum well, where $z\bot$2D plane. In this way we achieve independent, continuous control of the Rashba and Dresselhaus terms by using top and back gate voltages. This is an {\it unprecedented} tunability of the SO terms within a single sample.



{\it Detection scheme for the matched SO couplings.} WAL is a well established signature of SO coupling in magnetoconductance $\sigma(B_Z)$~\cite{bergmannreview, AAreview, iordanskii:1994, pikus:1995, knap:1996, miller:2003} exhibiting a local maximum at zero field. In the $|\alpha|=\beta$ regime, the resulting internal SO field is uniaxial,
spin rotations commute and are undone along time-reversal loops. Therefore WAL is suppressed and the effectively spin-less situation displaying weak localization (WL) (i.e., $\sigma(B_Z)$ exhibiting a local minimum at $B_Z=0$) is restored~\cite{pikus:1995,schliemann:2003,bernevig:2006,kohda:2012}.
Away from the matched regime, the SO field is not uniaxial, spin rotations do not commute and trajectories in time-reversal loops interfere destructively upon averaging~\cite{bergmannreview} due to the SO phases picked up along the loops thus leading to WAL. Hence this suppression is a sensitive detector for $\beta=\pm\alpha$. We note that the WL dip -- often used to determine phase coherence -- sensitively depends on the SO coupling (e.g. curves $3-6$ in Fig.\,\ref{fig:1}b), even before WAL appears. Negligence of SO coupling could thus lead to spurious or saturating coherence times.


{\it Continuous locking $\alpha=\beta$}. We proceed to demonstrate gate-locking of the SO couplings $\alpha$, $\beta$. Figure\,\ref{fig:1}b displays $\sigma(B_Z)$ of the $9.3\,$nm well for top and back gate configurations labeled $1-7$, all lying on a contour of constant density, see Fig.\,\ref{fig:1}a. Along this contour, $\beta$ is held fixed since the density is constant ($\beta_1$ is essentially gate independent), while $\alpha$ is changing as the gate voltages are modifying the electric field $\delta E_Z$ perpendicular to the quantum well. Across these gate configurations, the conductance shows a transition from WAL (conf. $1\,\&\,2$) to WL ($4\,\&\,5$) back to WAL ($7$). Selecting the most pronounced WL curve allows us to determine the symmetry point $\alpha=\beta$. This scheme is repeated for a number of densities, varying $n$ by a factor of $2$, yielding the symmetry point $\alpha=\beta$ for each density $n$ (see Fig.\,\ref{fig:2}a, blue markers), thus defining a symmetry line in the $(V_T,V_B)$-plane. Along this line, $\beta$ is changing with density as previously described, and $\alpha$ follows $\beta$, remaining ``continuously'' locked at $\alpha=\beta$. As mentioned earlier, this is a very interesting finding as it should allow the creation of persistent spin helices with gate-controllable pitches as illustrated in Fig.~\ref{fig:0}.

\begin{figure}\vspace{-3mm}
\includegraphics[width=8.7cm]{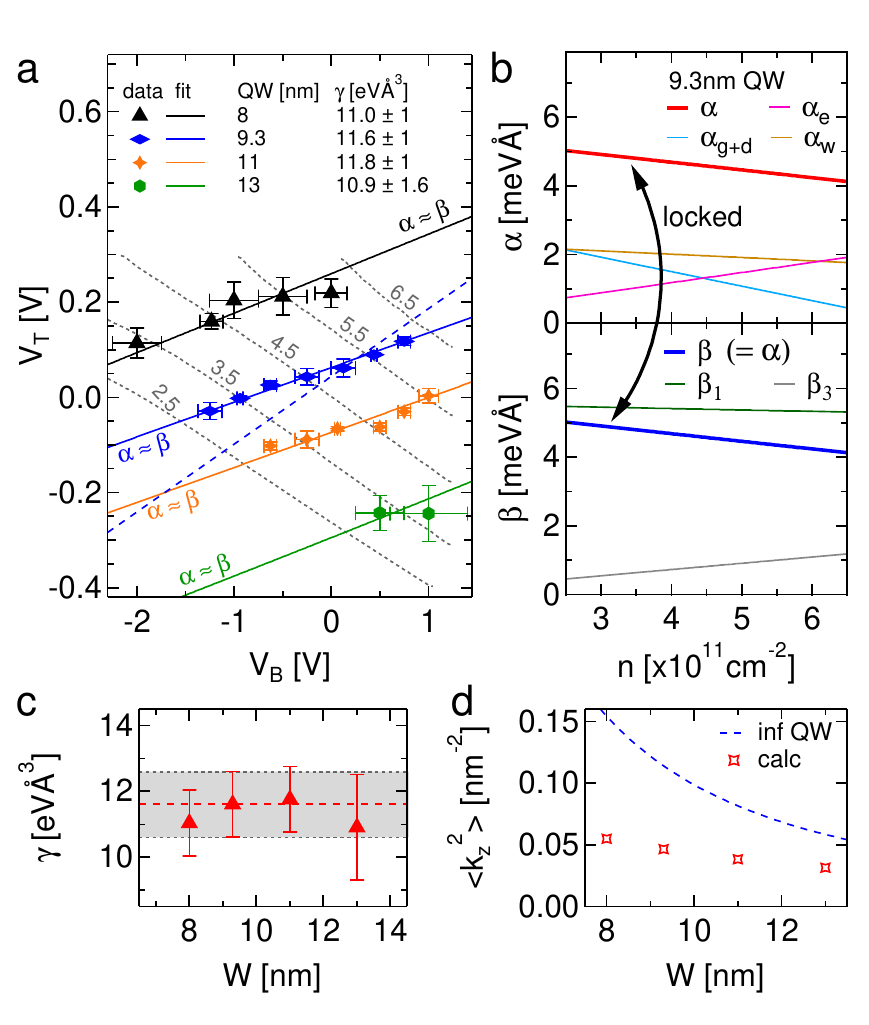}
\vspace{-7mm}
\caption{Tuning and continuously locking $\alpha=\beta$.
{\bf (a)} The markers indicate $\alpha\approx\beta$ for four different well widths (asymmetric doping) and various densities (gray contours of constant $n$, labeled in units of $10^{11}\,\mathrm{cm^{-2}}$) in the $V_T$ and $V_B$ plane. Error bars result from the finite number of conductance traces in the $(V_B,V_T)$-space. Theory fits (solid lines) are shown for each well, with $\gamma$ as the only fit parameter (inset table, error bars dominated by systematic error, see below). The dashed blue line indicates the slope of constant $\alpha=\beta_1$, neglecting $\beta_3$, which is inconsistent with the data. {\bf (b)} Simulation of locked $\alpha=\beta$ versus density $n$ along solid blue line from {\bf (a)}, showing the various SO contributions (see text). {\bf (c)} Values of $\gamma$ from fits for each well width $W$. Red dashed line is the average $\gamma=11.6\pm1\,\mathrm{eV\mathring{A}^3}$ (excluding $W=13\,\mathrm{nm}$ due to its larger error), gray the $\sim9\%$ error, stemming mostly from the systematic uncertainty in the input parameters of the simulations (see methods).  {\bf (d)} $\langle k_z^2\rangle$ as a function of well width $W$ for realistic (markers) and infinite (blue) potential.}\label{fig:2}\vspace{-6mm}
\end{figure}

{\it Simulations and fitting of $\gamma$.} Self-consistent calculations combined with the transport data can deliver all SO parameters. The numerical simulations~\cite{calsaverini:2008} (see Appendix and SOM) can accurately calculate $\alpha$ and $\langle k_z^2 \rangle$. This leaves only one fit parameter: $\gamma$, the bulk Dresselhaus coefficient, which can now be extracted from fits to the density dependence of the symmetry point, see solid blue line in Fig.\,\ref{fig:2}a, giving excellent agreement with the data (blue markers). This procedure can be repeated for a set of wafers with varying quantum well width and thus varying $\beta_1$. This shifts the symmetry point $\alpha=\beta$, producing nearly parallel lines, as indicated with colors in Fig.\,\ref{fig:2}a corresponding to the various wafers as labeled. As seen, locking $\alpha=\beta$ over a broad range is achieved in all wafers. Since gate voltages can be tuned continuously, any and all points on the symmetry lines $\alpha=\beta$ can be reached.  Again performing fits over the density dependence of the symmetry point for each well width, we obtain very good agreement, see Fig.\,\ref{fig:2}a, and extract $\gamma=11.6\pm1\,\mathrm{eV\mathring{A}^3}$ consistently for all wells (Fig.\,\ref{fig:2}c). We emphasize that $\gamma$ is notoriously difficult to calculate and measure~\cite{knap:1996,miller:2003,krich:2007}; the value reported here agrees well with recent studies~\cite{krich:2007,walser:2012_1, walser:2012_2}. Obtaining consistent values over wide ranges of densities and several wafers provides a robust method to extract $\gamma$.


Beyond $\gamma$, the simulations reveal important information about the gate-tuning of the SO parameters. The Rashba coefficient is modeled as $\alpha=\alpha_{\rm g+d}+\alpha_{\rm w}+\alpha_{\rm e}$ in the simulation, with gate and doping term $\alpha_{\rm g+d}$, quantum well structure term $\alpha_{\rm w}$, and Hartree term $\alpha_{\rm e}$. Along a contour of constant density, the simulations show that mainly $\alpha_{\rm g+d}$ and $\alpha_{\rm w}$ are modified, while $\alpha_{\rm e}$ and $\beta$ remain constant, see Fig.\,\ref{fig:1}c. The density dependence for locked $\alpha=\beta$, on the other hand, shows that while $\beta_1$ is nearly constant, $\beta_3$ is linearly increasing with $n$, thus reducing $\beta=\beta_1-\beta_3$, see Fig.\,\ref{fig:2}b. Hence, to keep $\alpha=\beta$ locked, $\alpha$ has to be reduced correspondingly. The Hartree term $\alpha_{\rm e}$, however, increases for growing $n$. Thus, on the $\alpha=\beta$ line, the other $\alpha$-terms -- mainly the gate dependent $\alpha_{\rm g+d}$ -- are strongly reduced, maintaining locked $\alpha=\beta$, as seen in Fig.\,\ref{fig:2}b. We emphasize that neglecting the gate/density dependence of $\beta_3$ and fixing $\alpha=\beta_1+\mathrm{const.}$ results in a line with slope indicated by the blue dashed line in Fig.\,\ref{fig:2}(a), which is clearly inconsistent with the data. Thus, the density dependent $\beta_3$ enabling gate-tunability of the Dresselhaus term is crucial here.


\begin{figure}\vspace{-0mm}
\includegraphics[width=8.7cm]{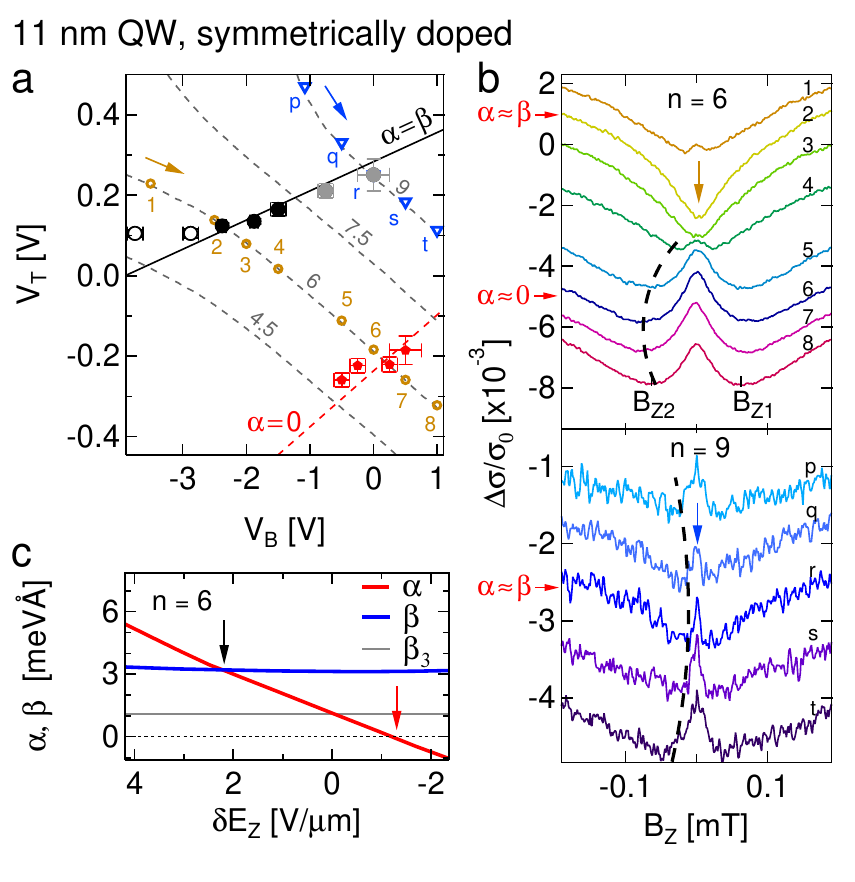}
\vspace{-7mm}\caption{The Dresselhaus and the cubic regime. {\bf (a)} Locked regime $\alpha\approx\beta$ (black/grey symbols) and Dresselhaus regime $\alpha\approx0$ (red symbols) from the broadest WAL minima (maximal $B_{\rm SO}$) in the $V_T$ and $V_B$ plane for a more symmetrically doped $11$\,nm well. The solid black line displays the $\alpha=\beta$ simulation, while the dashed red line marks the simulated $\alpha=0$ contour. Open black markers (leftmost $V_B$) are entering the non-linear gate regime, causing a slight deviation from theory, which assumes linear gate action. The rightmost $V_B$ points (gray) are obtained from the minimal $B_{\rm SO}$ in presence of WAL. {\bf (b)} Sequence at $n=6\cdot10^{11}\,\mathrm{cm^{-2}}$ (upper panel) and $n=9\cdot10^{11}\,\mathrm{cm^{-2}}$ (lower panel), shifted vertically for clarity. Each brown/blue marker in {\bf (a)} corresponds to a trace in {\bf (b)}, as labeled by numerals/letters. $B_{\rm SO}$ is indicated as a guide for the eye by black dashed curves for negative $B_Z$. $B_{\rm SO}$ increases and peaks (indicating $\alpha=0$) before decreasing again (upper panel). Broken spin symmetry regime (lower panel): WAL is no longer suppressed here due to symmetry breaking from the cubic term at large $n$. Still, $\alpha\approx\beta$ can be identified with the narrowest WAL peak.  {\bf (c)} Simulation of $\alpha$ and $\beta$ along $n=6\cdot10^{11}\,\mathrm{cm^{-2}}$. $\alpha$ traverses both $\beta$ (black arrow) and for smaller $\delta E_Z$ also zero (red arrow).
}\label{fig:3}\vspace{-5mm}
\end{figure}

{\it Dresselhaus regime.} We now show that $\alpha$ can be tuned through $\beta$ \emph{and through zero} in a more symmetrically doped wafer, opening the Dresselhaus regime $\beta\gg\alpha$. We introduce the magnetic field $B_{\rm SO}$ where the magneto conductance exhibits minima at $B_{\rm Z1}\approx-B_{\rm Z2}$. These minima describe the crossover between WAL and WL, where the Aharonov-Bohm dephasing length  and the  SO diffusion length are comparable.  Beyond the WAL-WL-WAL transition (Fig.\,\ref{fig:3}b upper panel), $B_{\rm SO}$ is seen to peak and decrease again (dashed curve). The gate voltages with maximal $B_{\rm SO}$ are added to Fig.\,\ref{fig:3}a for several densities (red markers). We surmise that these points mark $\alpha\approx0$: $B_{\rm SO}$ signifies the crossover between WL/WAL-like conductance, thus defining an empirical measure for the effects of SO coupling (larger $B_{\rm SO}$, stronger effects). For $\alpha=0$, the full effect of $\beta$ on the conductance becomes apparent without cancellation from $\alpha$, giving a maximal $B_{\rm SO}$. Indeed, the simulated $\alpha=0$ curve (dashed red line in Fig.\,\ref{fig:3}a) cuts through the experimental points, also reflected in Fig.\,\ref{fig:3}c by a good match with the simulated $\alpha=0$ crossing point (red arrow).

{\it Diverging spin-orbit lengths.} For a comparison of experiment and simulation, we convert the empirical $B_{\rm SO}$ to a ``magnetic length'' $\lambda_{\rm SO}=\sqrt{\hbar/2eB_{\rm SO}}$, which we later on interpret as a spin-diffusion length, where $e>0$ is the electron charge and the factor of two accounts for time-reversed pairs of closed trajectories. We also introduce the {\it ballistic} SO lengths $\lambda_{\pm}=\hbar^2/(2m^*\left|\alpha\pm\beta\right|)$. These lengths correspond to a spin rotation of 1 radian, as the electrons travel along $\hat{x}_+$ and $\hat{x}_-$, respectively, with spins initially aligned perpendicular to the corresponding SO field (e.g., for an electron moving along the $\hat{x}_+$ its spin should point along $\hat{x}_+$ or $\hat{z}$ so spin precession can occur, see SOM Eq~S20 for an expression of the SO field). For $\beta=+\alpha$, $\lambda_-$ diverges (no precession, indicating that an electron traveling along $\hat x_-$ does not precess) while $\lambda_+$ is finite, and vice versa for $\beta=-\alpha$.

Figure~\ref{fig:4} shows the theoretical spin diffusion length $\lambda_{\mathrm{eff}}$ (see methods) and the ballistic $\lambda_\pm$, together with the experimental $\lambda_{\rm SO}$, all agreeing remarkably well. Since at $\alpha=\beta$ spin transport is ballistic despite charge diffusion, $\lambda_-$ and its diffusive counterpart $\lambda_{\rm eff}$ (small $\beta_3$) are essentially equivalent as shown in the SOM. The enhanced $\lambda_{\rm SO}$ around $\alpha/\beta=1$ corresponds to an increased spin relaxation time $\tau_{\rm SO}=\lambda_{\rm SO}^2/(2D)$. Note that $\max(\lambda_\pm)$ quantifies the deviation from the \emph{uniaxial} SO field away from $\alpha=\beta$, and thus the extent to which spin rotations are not undone in a closed trajectory due to the non-Abelian nature of spin rotations around non-collinear axes. This leads to WAL, a finite $B_{\rm SO}$ and $\lambda_{\rm SO}\simeq \max(\lambda_\pm)$, as observed (see Fig.\,\ref{fig:4}). Unlike the corresponding time scales, the SO lengths are only weakly dependent on density and mobility when plotted against $\alpha/\beta$, allowing a comparison of various densities.

\begin{figure}[t]
\vspace{-5mm}\includegraphics[width=8.7cm]{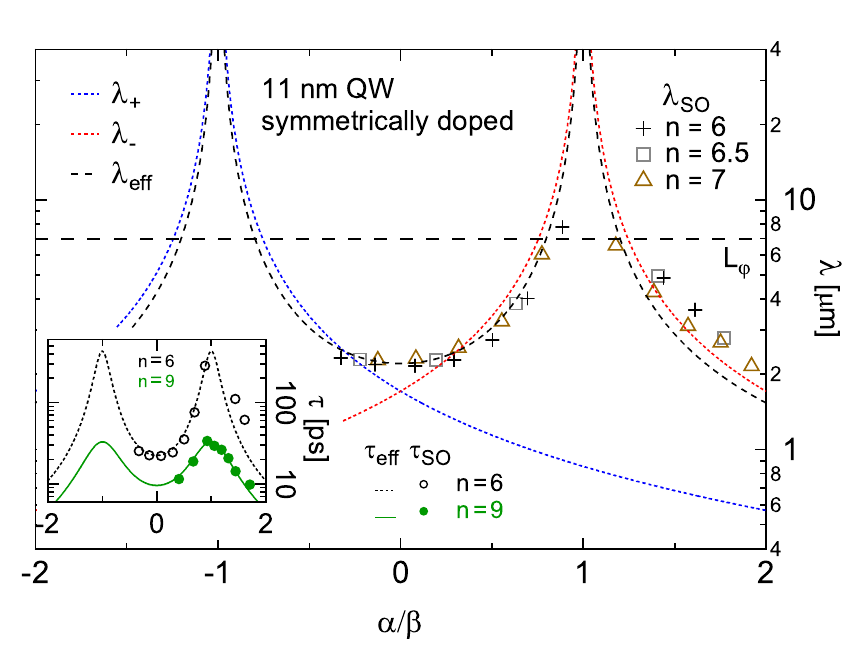}
\vspace{-3mm}\caption{
Experimental and theoretical SO-lengths and SO-times. Experimental $\lambda_{\rm SO}=\sqrt{\hbar/2eB_{\rm SO}}$ (markers, densities as labeled, in units of $10^{11}\,\mathrm{cm^{-2}}$) as a function of the dimensionless ratio $\alpha/\beta$ (from SO simulation). The ballistic $\lambda_{\pm}$ (blue/red dashes) and effective $\lambda_{\mathrm{eff}}$ (black dashed curve) are only weakly $n$-dependent (small $\beta_3$) when plotted against $\alpha/\beta$. Thus, curves for only one density ($n=6\cdot10^{11}\,\mathrm{cm^{-2}}$) are shown. The experimental uncertainty on $\lambda_{\rm SO}$ is captured by the spread given by the three slightly different densities. The coherence length $L_\varphi\approx7\,\mathrm{\mu m}$ is added for illustration (obtained from WL curves), setting the visibility of SO effects on the conductance and thus the width of the WAL-WL-WAL transition. Inset: experimental spin relaxation time $\tau_{\rm SO}=\lambda^2_{\rm SO}/(2D)$ (circles) as a function of $\alpha/\beta$ for two densities as indicated. Theory curves $\tau_{\mathrm{eff}}$ (dashed) now include the symmetry breaking third harmonic term, preventing divergence at $\alpha/\beta=1$, while $\lambda_{\rm eff}$ (main panel) does not.}
\label{fig:4}\vspace{-5mm}
\end{figure}

The third harmonic contribution of cubic-in-$k$ term causes spin relaxation even at $\alpha=\beta$ and becomes visible at large densities: WAL is present in all traces and through $\alpha=\beta$ (Fig.\,\ref{fig:3}b, lower panel), because the SO field can no longer be made uniaxial, thus breaking spin symmetry and reviving WAL. A partial symmetry restoration is still apparent, where -- in contrast to the $\alpha=0$ case -- a \emph{minimal} $B_{\rm SO}$ is reached (dashed curves) consistent with $\alpha=\beta$ (grey markers Fig.\,\ref{fig:3}a at large $n$). We include the cubic $\beta_3$ in the spin relaxation time $\tau_{\mathrm{eff}}$ (see methods), shown in the inset of Fig.\,\ref{fig:4} for two densities, finding good agreement with the experimental $\tau_{\rm SO}=\lambda^2_{\rm SO}/(2D)$, where $D$ is the diffusion constant. Over the whole locked regime of Fig.\,\ref{fig:2}b, WAL is absent, and $\tau_{\rm SO}$ is enhanced between one and two orders of magnitude compared to $\alpha=0$. Finally, the coherence length $L_\varphi$ sets an upper limit for the visibility of SO effects: WAL is suppressed for $\lambda_{\rm eff}\gg L_\varphi$, setting the width of the WAL-WL-WAL transition (see SOM).

{\it Final remarks and outlook.---} This work is laying the foundation for a new generation of experiments benefiting from unprecedented command over SO coupling in semiconductor nanostructures such as quantum wires, quantum dots, and electron spin qubits. Moreover, our work relaxes the stringency (i.e., the ``fine tuning") of the $\alpha=\beta$ symmetry condition at a particular singular point (gate) by introducing a ``continuous locking''  of the SO couplings $\alpha(V_T,V_B)=\beta(V_T,V_B)$ over a wide range of voltages, which should enable new experiments exciting persistent spin helices with variable pitches in GaAs wells~\cite{koralek:2009,walser:2012_1}, i.e., {\it stretchable PSHs}.  Another possibility is the generation of a skyrmion lattice (crossed spin helices) with variable lattice constants, as recently proposed in Ref.~\cite{fu:2015}.

Finally, we stress that within the continuously-locked regime of SO couplings demonstrated in our study, SO-coupled quantum transport in our samples shows a very distinctive feature: it is diffusive (2D) for charge while ballistic (1D) for spins thus providing a unique setting for coherent spin control. This ultimately adds a new functionality to the non-ballistic spin transistor of Ref.~\cite{schliemann:2003}, i.e. it can now be made to operate as the ideal (ballistic) Datta-Das spin transistor -- but in a realistic 2D diffusive system, with yet controlled spin rotations protected from spin decay.


\acknowledgements
We would like to thank A. C. Gossard, D. Loss, D. L. Maslov, G. Salis for valuable inputs and stimulating discussions. This work was supported by the Swiss Nanoscience Institute (SNI), NCCR QSIT, Swiss NSF, ERC starting grant, EU-FP7 SOLID and MICROKELVIN, US NSF and ONR, Brazilian grants FAPESP, CNPq, PRP/USP (Q-NANO), and natural science foundation of China (Grant No.~11004120).

\section*{Author Contributions}
F.D., J.F., P.W., J.C.E. and D.M.Z. designed the experiments, analysed the data and co-wrote the paper. All authors discussed the results and commented on the manuscript. S.M. and D.D.A. designed, simulated, and carried out the molecular beam epitaxy growth of the heterostructures. F.D. processed the samples and with P.W. performed the experiments. J.F. and J.C.E. developed and carried out the simulations and theoretical work.

\appendix*
\section{Materials and Methods}
\subsection{GaAs quantum well materials}
The wells are grown on an n-doped substrate (for details see SOM) and fabricated into Hall bar structures (see inset, Fig.\,\ref{fig:1}a) using standard photolithographic methods. The 2D gas is contacted by thermally annealed GeAu/Pt Ohmic contacts, optimized for a low contact resistance while maintaining high back gate tunability (low leakage currents) and avoiding short circuits to the back gate. On one segment of the Hall bar, a Ti/Au top gate with dimensions of $300$\,x\,$100$\,$\mu m^2$ was deposited. The average gate-induced E-field change in the well is defined as $\delta E_Z=1/2\left(V_T/d_T-V_B/d_B\right)$, with effective distance $d_{T/B}$ from the well to the top/back gate, respectively, extracted using a capacitor model, consistent with the full quantum description (see SOM). Contours of constant density follow $\delta V_T/d_T=-\delta V_B/d_B$. Deviations from linear behavior appear at most positive/negative gate voltages due to incipient gate leakage and hysteresis.

\subsection{Low temperature electronic measurements.}
The experiments are performed in a dilution refrigerator with base temperature $20$\,mK. We have used a standard four-wire lock-in technique at $133\,$Hz and $100\,$nA current bias, chosen to avoid self-heating while maximising the signal. The density is determined with Hall measurements in the classical regime, whereas Shubnikov-de Haas oscillations were used to exclude occupation of the second subband, which is the case for all the data discussed. The WAL signature is a small correction ($10^{-3}$) to total conductance. To achieve a satisfactory signal-to-noise ratio, longitudinal conductivity traces $\Delta\sigma/\sigma_0=(\sigma(B)-\sigma(0))/\sigma(0)$ were measured at least $10$ times and averaged.

\subsection{\bf Numerical Simulations}
The simulations calculate the Rashba coefficient $\alpha$ and $\langle k_z^2\rangle$ based on the bulk semiconductor band parameters, the well structure, the measured  electron densities and the measured gate lever arms. We solve the Schr\"odinger and Poisson equations self consistently (``Hartree approximation''), obtain the self-consistent eigenfunctions, and then determine $\alpha$ via appropriate expectation values~\cite{calsaverini:2008}. The Dresselhaus coefficient $\gamma$ is extracted from fits of the simulation to the experiment which detects the absence of WAL at $\alpha=\beta=\gamma(\langle k_z^2\rangle-k_F^2/4)$. Thus, given $\alpha$ and $\langle k_z^2 \rangle$ from the simulation and the measured $n=k_F^2/(2\pi)$, we obtain $\gamma=11.6\pm1\,\mathrm{eV\AA^3}$ consistently for all asymmetrically doped wells. Taking into account the uncertainties of the band parameters, the experimental errors and a negligible uncertainty on $\langle k_z^2 \rangle$, an overall uncertainty of about $9-10\%$ or about $\pm 1\,\mathrm{eV\AA^3}$ on $\gamma$ results. About $1-2\%$ error originates from the experimental uncertainty of determining $\alpha=\beta$. The doping distribution (above/below well) is not expected to influence $\gamma$, and hence we use the same $\gamma$ for the more symmetrically doped wafer. Fits to the $\alpha=\beta$ experimental points then determine how much charge effectively comes from upper rather than lower doping layers, fixing the last unknown parameter also for the more symmetrically doped well (see SOM).

\subsection{Spin-dephasing times and lengths}
In WL/WAL measurements, additional spin dephasing is introduced by the external magnetic field $B$ via the Aharonov-Bohm phase arising from the magnetic flux enclosed by the time reversed trajectories: $\Delta \varphi =2eAB/\hbar$, where $A$ is the loop area. Here we take $A=\lambda_{\rm SO}^2=2D\tau_{\rm SO}$ as a characteristic ``diffusion area'' probed by our WL/WAL experiment, with $\tau_{\rm SO}$ being the spin dephasing time, and $\lambda_{\rm SO}$ the spin diffusion length. By taking $\Delta \varphi=1$ (rad) at $B = B_{\rm SO}$, we can extract the spin-diffusion length $\lambda_{\rm SO}$ and spin-dephasing time $\tau_{\rm SO}$ from the minima of the WAL curves from $\lambda_{\rm SO}=\sqrt{\hbar/2eB_{\rm SO}}$ and $\tau_{\rm SO} = \hbar (4eDB_{\rm SO})^{-1}$, respectively. The factor of 4 here stems from the two time-reversed paths and the diffusion length.

\subsection{Effective SO times and lengths}
Theoretically, we determine $\tau_{\rm SO}$ via a spin random walk process (D'yakonov-Perel (DP)). The initial electron spin in a loop can point (with equal probability) along the $s_{x_-}$, $s_{x_+}$, and $s_z$ axes (analogous to $x_+$, $x_-$, and $z$, respectively), which have unequal spin-dephasing times $\tau_{\rm DP,s_{x_-}}$, $\tau_{\rm DP,s_{x_+}}$, and $\tau_{\rm DP,s_z}$. For unpolarized, independent spins, we take  the average $\tau_{\rm eff} =(\tau_{\rm DP,s_{x_-}}+\tau_{\rm DP,s_{x_+}}+\tau_{\rm DP,s_z})/3$, which leads to an effective spin difusion length $\lambda_{\rm eff}= \sqrt{2D \tau_{\rm eff}}$. Actually, $\lambda_{\rm eff}$ is defined from the average variance $\lambda^2_{\rm eff}=\bar \sigma^2=2D\tau_{\rm eff}$, obtained by averaging the spin-dependent variances $\sigma^2_{s_{x_-}}=2D\tau_{\rm DP,s_{x_-}}$, $\sigma^2_{s_{x_+}}=2D\tau_{\rm DP,s_{x_+}}$ and $\sigma^2_{s_{z}}=2D\tau_{\rm DP,s_{z}}$ over the spin directions $s_{x_+}$, $s_{x_-}$, and $s_z$ (this is equivalent to averaging over the $\tau$'s and not over $1/\tau$'s). In the SOM, we discuss the spin random walk and provide expressions for the DP times including corrections due to the cubic $\beta_3$ term. Figure~\ref{fig:4} shows curves for the spin dephasing times and lengths presented here. In the main panel, the cubic $\beta_3$ is neglected in $\lambda_{\rm eff}$ since for $n\leq7\,\cdot10^{11}\,\mathrm{cm^{-2}}$, WL appears at $\alpha=\beta$ (small $\beta_3$). In contrast, the cubic term is included in $\tau_{\rm eff}$ in the inset since at the higher density $n=9\,\cdot10^{11}\,\mathrm{cm^{-2}}$, WAL persists (sufficiently strong $\beta_3$).


\begin{thebibliography}{27}%
\makeatletter
\providecommand \@ifxundefined [1]{%
 \@ifx{#1\undefined}
}%
\providecommand \@ifnum [1]{%
 \ifnum #1\expandafter \@firstoftwo
 \else \expandafter \@secondoftwo
 \fi
}%
\providecommand \@ifx [1]{%
 \ifx #1\expandafter \@firstoftwo
 \else \expandafter \@secondoftwo
 \fi
}%
\providecommand \natexlab [1]{#1}%
\providecommand \enquote  [1]{``#1''}%
\providecommand \bibnamefont  [1]{#1}%
\providecommand \bibfnamefont [1]{#1}%
\providecommand \citenamefont [1]{#1}%
\providecommand \href@noop [0]{\@secondoftwo}%
\providecommand \href [0]{\begingroup \@sanitize@url \@href}%
\providecommand \@href[1]{\@@startlink{#1}\@@href}%
\providecommand \@@href[1]{\endgroup#1\@@endlink}%
\providecommand \@sanitize@url [0]{\catcode `\\12\catcode `\$12\catcode
  `\&12\catcode `\#12\catcode `\^12\catcode `\_12\catcode `\%12\relax}%
\providecommand \@@startlink[1]{}%
\providecommand \@@endlink[0]{}%
\providecommand \url  [0]{\begingroup\@sanitize@url \@url }%
\providecommand \@url [1]{\endgroup\@href {#1}{\urlprefix }}%
\providecommand \urlprefix  [0]{URL }%
\providecommand \Eprint [0]{\href }%
\providecommand \doibase [0]{http://dx.doi.org/}%
\providecommand \selectlanguage [0]{\@gobble}%
\providecommand \bibinfo  [0]{\@secondoftwo}%
\providecommand \bibfield  [0]{\@secondoftwo}%
\providecommand \translation [1]{[#1]}%
\providecommand \BibitemOpen [0]{}%
\providecommand \bibitemStop [0]{}%
\providecommand \bibitemNoStop [0]{.\EOS\space}%
\providecommand \EOS [0]{\spacefactor3000\relax}%
\providecommand \BibitemShut  [1]{\csname bibitem#1\endcsname}%
\let\auto@bib@innerbib\@empty
\bibitem [{\citenamefont {Sinova}\ \emph {et~al.}(2015)\citenamefont {Sinova},
  \citenamefont {Valenzuela}, \citenamefont {Wunderlich}, \citenamefont
  {Back},\ and\ \citenamefont {Jungwirth}}]{sinova:2015}%
  \BibitemOpen
  \bibfield  {author} {\bibinfo {author} {\bibfnamefont {J.}~\bibnamefont
  {Sinova}}, \bibinfo {author} {\bibfnamefont {S.~O.}\ \bibnamefont
  {Valenzuela}}, \bibinfo {author} {\bibfnamefont {J.}~\bibnamefont
  {Wunderlich}}, \bibinfo {author} {\bibfnamefont {C.}~\bibnamefont {Back}}, \
  and\ \bibinfo {author} {\bibfnamefont {T.}~\bibnamefont {Jungwirth}},\ }\href
  {\doibase 10.1103/RevModPhys.87.1213} {\bibfield  {journal} {\bibinfo
  {journal} {Reviews of Modern Physics}\ }\textbf {\bibinfo {volume} {87}},\
  \bibinfo {pages} {1213} (\bibinfo {year} {2015})}\BibitemShut {NoStop}%
\bibitem [{\citenamefont {Mourik}\ \emph {et~al.}(2012)\citenamefont {Mourik},
  \citenamefont {Zuo}, \citenamefont {Frolov}, \citenamefont {Plissard},
  \citenamefont {Bakkers},\ and\ \citenamefont {Kouwenhoven}}]{mourik:2012}%
  \BibitemOpen
  \bibfield  {author} {\bibinfo {author} {\bibfnamefont {V.}~\bibnamefont
  {Mourik}}, \bibinfo {author} {\bibfnamefont {K.}~\bibnamefont {Zuo}},
  \bibinfo {author} {\bibfnamefont {S.~M.}\ \bibnamefont {Frolov}}, \bibinfo
  {author} {\bibfnamefont {S.~R.}\ \bibnamefont {Plissard}}, \bibinfo {author}
  {\bibfnamefont {E.~P. a.~M.}\ \bibnamefont {Bakkers}}, \ and\ \bibinfo
  {author} {\bibfnamefont {L.~P.}\ \bibnamefont {Kouwenhoven}},\ }\href
  {\doibase 10.1126/science.1222360} {\bibfield  {journal} {\bibinfo  {journal}
  {Science}\ }\textbf {\bibinfo {volume} {336}},\ \bibinfo {pages} {1003}
  (\bibinfo {year} {2012})}\BibitemShut {NoStop}%
\bibitem [{\citenamefont {Wan}\ \emph {et~al.}(2011)\citenamefont {Wan},
  \citenamefont {Turner}, \citenamefont {Vishwanath},\ and\ \citenamefont
  {Savrasov}}]{wan:2011}%
  \BibitemOpen
  \bibfield  {author} {\bibinfo {author} {\bibfnamefont {X.}~\bibnamefont
  {Wan}}, \bibinfo {author} {\bibfnamefont {A.~M.}\ \bibnamefont {Turner}},
  \bibinfo {author} {\bibfnamefont {A.}~\bibnamefont {Vishwanath}}, \ and\
  \bibinfo {author} {\bibfnamefont {S.~Y.}\ \bibnamefont {Savrasov}},\ }\href
  {\doibase 10.1103/PhysRevB.83.205101} {\bibfield  {journal} {\bibinfo
  {journal} {Physical Review B}\ }\textbf {\bibinfo {volume} {83}},\ \bibinfo
  {pages} {205101} (\bibinfo {year} {2011})}\BibitemShut {NoStop}%
\bibitem [{\citenamefont {Engels}\ \emph {et~al.}(1997)\citenamefont {Engels},
  \citenamefont {Lange}, \citenamefont {Sch\"apers},\ and\ \citenamefont
  {L\"uth}}]{engels:1997}%
  \BibitemOpen
  \bibfield  {author} {\bibinfo {author} {\bibfnamefont {G.}~\bibnamefont
  {Engels}}, \bibinfo {author} {\bibfnamefont {J.}~\bibnamefont {Lange}},
  \bibinfo {author} {\bibfnamefont {T.}~\bibnamefont {Sch\"apers}}, \ and\
  \bibinfo {author} {\bibfnamefont {H.}~\bibnamefont {L\"uth}},\ }\href@noop {}
  {\bibfield  {journal} {\bibinfo  {journal} {Physical Review B}\ }\textbf
  {\bibinfo {volume} {55}},\ \bibinfo {pages} {R1958} (\bibinfo {year}
  {1997})}\BibitemShut {NoStop}%
\bibitem [{\citenamefont {Nitta}\ \emph {et~al.}(1997)\citenamefont {Nitta},
  \citenamefont {Akazaki}, \citenamefont {Takayanagi},\ and\ \citenamefont
  {Enoki}}]{nitta:1997}%
  \BibitemOpen
  \bibfield  {author} {\bibinfo {author} {\bibfnamefont {J.}~\bibnamefont
  {Nitta}}, \bibinfo {author} {\bibfnamefont {T.}~\bibnamefont {Akazaki}},
  \bibinfo {author} {\bibfnamefont {H.}~\bibnamefont {Takayanagi}}, \ and\
  \bibinfo {author} {\bibfnamefont {T.}~\bibnamefont {Enoki}},\ }\href@noop {}
  {\bibfield  {journal} {\bibinfo  {journal} {Physical Review Letters}\
  }\textbf {\bibinfo {volume} {78}},\ \bibinfo {pages} {1335} (\bibinfo {year}
  {1997})}\BibitemShut {NoStop}%
\bibitem [{\citenamefont {Chuang}\ \emph {et~al.}(2015)\citenamefont {Chuang},
  \citenamefont {Ho}, \citenamefont {Smith}, \citenamefont {Sfigakis},
  \citenamefont {Pepper}, \citenamefont {Chen}, \citenamefont {Fan},
  \citenamefont {Griffiths}, \citenamefont {Farrer}, \citenamefont {Beere},
  \citenamefont {Jones}, \citenamefont {Ritchie},\ and\ \citenamefont
  {Chen}}]{chuang:2015}%
  \BibitemOpen
  \bibfield  {author} {\bibinfo {author} {\bibfnamefont {P.}~\bibnamefont
  {Chuang}}, \bibinfo {author} {\bibfnamefont {S.-C.}\ \bibnamefont {Ho}},
  \bibinfo {author} {\bibfnamefont {L.~W.}\ \bibnamefont {Smith}}, \bibinfo
  {author} {\bibfnamefont {F.}~\bibnamefont {Sfigakis}}, \bibinfo {author}
  {\bibfnamefont {M.}~\bibnamefont {Pepper}}, \bibinfo {author} {\bibfnamefont
  {C.-H.}\ \bibnamefont {Chen}}, \bibinfo {author} {\bibfnamefont {J.-C.}\
  \bibnamefont {Fan}}, \bibinfo {author} {\bibfnamefont {J.~P.}\ \bibnamefont
  {Griffiths}}, \bibinfo {author} {\bibfnamefont {I.}~\bibnamefont {Farrer}},
  \bibinfo {author} {\bibfnamefont {H.~E.}\ \bibnamefont {Beere}}, \bibinfo
  {author} {\bibfnamefont {G.~a.~C.}\ \bibnamefont {Jones}}, \bibinfo {author}
  {\bibfnamefont {D.~A.}\ \bibnamefont {Ritchie}}, \ and\ \bibinfo {author}
  {\bibfnamefont {T.-M.}\ \bibnamefont {Chen}},\ }\href {\doibase
  10.1038/nnano.2014.296} {\bibfield  {journal} {\bibinfo  {journal} {Nature
  Nanotechnology}\ }\textbf {\bibinfo {volume} {10}},\ \bibinfo {pages} {35}
  (\bibinfo {year} {2015})}\BibitemShut {NoStop}%
\bibitem [{\citenamefont {Datta}\ and\ \citenamefont {Das}(1990)}]{datta:1990}%
  \BibitemOpen
  \bibfield  {author} {\bibinfo {author} {\bibfnamefont {S.}~\bibnamefont
  {Datta}}\ and\ \bibinfo {author} {\bibfnamefont {B.}~\bibnamefont {Das}},\
  }\href@noop {} {\bibfield  {journal} {\bibinfo  {journal} {Applied Physics
  Letters}\ }\textbf {\bibinfo {volume} {56}},\ \bibinfo {pages} {665}
  (\bibinfo {year} {1990})}\BibitemShut {NoStop}%
\bibitem [{\citenamefont {Bychkov}\ and\ \citenamefont
  {Rashba}(1984)}]{rashba:1984}%
  \BibitemOpen
  \bibfield  {author} {\bibinfo {author} {\bibfnamefont {Y.~A.}\ \bibnamefont
  {Bychkov}}\ and\ \bibinfo {author} {\bibfnamefont {E.~I.}\ \bibnamefont
  {Rashba}},\ }\href@noop {} {\bibfield  {journal} {\bibinfo  {journal} {{JETP}
  Letters}\ }\textbf {\bibinfo {volume} {39}},\ \bibinfo {pages} {78} (\bibinfo
  {year} {1984})}\BibitemShut {NoStop}%
\bibitem [{\citenamefont {Dresselhaus}(1955)}]{dresselhaus:1955}%
  \BibitemOpen
  \bibfield  {author} {\bibinfo {author} {\bibfnamefont {G.}~\bibnamefont
  {Dresselhaus}},\ }\href@noop {} {\bibfield  {journal} {\bibinfo  {journal}
  {Physical Review}\ }\textbf {\bibinfo {volume} {100}},\ \bibinfo {pages}
  {580} (\bibinfo {year} {1955})}\BibitemShut {NoStop}%
\bibitem [{\citenamefont {Schliemann}\ \emph {et~al.}(2003)\citenamefont
  {Schliemann}, \citenamefont {Egues},\ and\ \citenamefont
  {Loss}}]{schliemann:2003}%
  \BibitemOpen
  \bibfield  {author} {\bibinfo {author} {\bibfnamefont {J.}~\bibnamefont
  {Schliemann}}, \bibinfo {author} {\bibfnamefont {J.~C.}\ \bibnamefont
  {Egues}}, \ and\ \bibinfo {author} {\bibfnamefont {D.}~\bibnamefont {Loss}},\
  }\href {\doibase 10.1103/PhysRevLett.90.146801} {\bibfield  {journal}
  {\bibinfo  {journal} {Physical Review Letters}\ }\textbf {\bibinfo {volume}
  {90}},\ \bibinfo {pages} {085323} (\bibinfo {year} {2003})}\BibitemShut
  {NoStop}%
\bibitem [{\citenamefont {Bernevig}\ \emph {et~al.}(2006)\citenamefont
  {Bernevig}, \citenamefont {Orenstein},\ and\ \citenamefont
  {Zhang}}]{bernevig:2006}%
  \BibitemOpen
  \bibfield  {author} {\bibinfo {author} {\bibfnamefont {B.}~\bibnamefont
  {Bernevig}}, \bibinfo {author} {\bibfnamefont {J.}~\bibnamefont {Orenstein}},
  \ and\ \bibinfo {author} {\bibfnamefont {S.-C.}\ \bibnamefont {Zhang}},\
  }\href {\doibase 10.1103/PhysRevLett.97.236601} {\bibfield  {journal}
  {\bibinfo  {journal} {Physical Review Letters}\ }\textbf {\bibinfo {volume}
  {97}},\ \bibinfo {pages} {236601} (\bibinfo {year} {2006})}\BibitemShut
  {NoStop}%
\bibitem [{\citenamefont {Koralek}\ \emph {et~al.}(2009)\citenamefont
  {Koralek}, \citenamefont {Weber}, \citenamefont {Orenstein}, \citenamefont
  {Bernevig}, \citenamefont {Zhang}, \citenamefont {Mack},\ and\ \citenamefont
  {Awschalom}}]{koralek:2009}%
  \BibitemOpen
  \bibfield  {author} {\bibinfo {author} {\bibfnamefont {J.~D.}\ \bibnamefont
  {Koralek}}, \bibinfo {author} {\bibfnamefont {C.~P.}\ \bibnamefont {Weber}},
  \bibinfo {author} {\bibfnamefont {J.}~\bibnamefont {Orenstein}}, \bibinfo
  {author} {\bibfnamefont {B.~A.}\ \bibnamefont {Bernevig}}, \bibinfo {author}
  {\bibfnamefont {S.-C.}\ \bibnamefont {Zhang}}, \bibinfo {author}
  {\bibfnamefont {S.}~\bibnamefont {Mack}}, \ and\ \bibinfo {author}
  {\bibfnamefont {D.~D.}\ \bibnamefont {Awschalom}},\ }\href {\doibase
  10.1038/nature07871} {\bibfield  {journal} {\bibinfo  {journal} {Nature}\
  }\textbf {\bibinfo {volume} {458}},\ \bibinfo {pages} {610} (\bibinfo {year}
  {2009})}\BibitemShut {NoStop}%
\bibitem [{\citenamefont {Walser}\ \emph
  {et~al.}(2012{\natexlab{a}})\citenamefont {Walser}, \citenamefont {Reichl},
  \citenamefont {Wegscheider},\ and\ \citenamefont {Salis}}]{walser:2012_1}%
  \BibitemOpen
  \bibfield  {author} {\bibinfo {author} {\bibfnamefont {M.~P.}\ \bibnamefont
  {Walser}}, \bibinfo {author} {\bibfnamefont {C.}~\bibnamefont {Reichl}},
  \bibinfo {author} {\bibfnamefont {W.}~\bibnamefont {Wegscheider}}, \ and\
  \bibinfo {author} {\bibfnamefont {G.}~\bibnamefont {Salis}},\ }\href
  {\doibase 10.1038/nphys2383} {\bibfield  {journal} {\bibinfo  {journal}
  {Nature Physics}\ }\textbf {\bibinfo {volume} {8}},\ \bibinfo {pages} {757}
  (\bibinfo {year} {2012}{\natexlab{a}})}\BibitemShut {NoStop}%
\bibitem [{\citenamefont {Kohda}\ \emph {et~al.}(2012)\citenamefont {Kohda},
  \citenamefont {Lechner}, \citenamefont {Kunihashi}, \citenamefont
  {Dollinger}, \citenamefont {Olbrich}, \citenamefont {Sch\"onhuber},
  \citenamefont {Caspers}, \citenamefont {Bel'kov}, \citenamefont {Golub},\
  and\ \citenamefont {Weiss}}]{kohda:2012}%
  \BibitemOpen
  \bibfield  {author} {\bibinfo {author} {\bibfnamefont {M.}~\bibnamefont
  {Kohda}}, \bibinfo {author} {\bibfnamefont {V.}~\bibnamefont {Lechner}},
  \bibinfo {author} {\bibfnamefont {Y.}~\bibnamefont {Kunihashi}}, \bibinfo
  {author} {\bibfnamefont {T.}~\bibnamefont {Dollinger}}, \bibinfo {author}
  {\bibfnamefont {P.}~\bibnamefont {Olbrich}}, \bibinfo {author} {\bibfnamefont
  {C.}~\bibnamefont {Sch\"onhuber}}, \bibinfo {author} {\bibfnamefont
  {I.}~\bibnamefont {Caspers}}, \bibinfo {author} {\bibfnamefont {V.~V.}\
  \bibnamefont {Bel'kov}}, \bibinfo {author} {\bibfnamefont {L.~E.}\
  \bibnamefont {Golub}}, \ and\ \bibinfo {author} {\bibfnamefont
  {D.}~\bibnamefont {Weiss}},\ }\href@noop {} {\bibfield  {journal} {\bibinfo
  {journal} {Physical Review B}\ }\textbf {\bibinfo {volume} {86}},\ \bibinfo
  {pages} {081306} (\bibinfo {year} {2012})}\BibitemShut {NoStop}%
\bibitem [{\citenamefont {Iordanskii}\ \emph {et~al.}(1994)\citenamefont
  {Iordanskii}, \citenamefont {Lyanda-Geller},\ and\ \citenamefont
  {Pikus}}]{iordanskii:1994}%
  \BibitemOpen
  \bibfield  {author} {\bibinfo {author} {\bibfnamefont {S.~V.}\ \bibnamefont
  {Iordanskii}}, \bibinfo {author} {\bibfnamefont {Y.~B.}\ \bibnamefont
  {Lyanda-Geller}}, \ and\ \bibinfo {author} {\bibfnamefont {G.~E.}\
  \bibnamefont {Pikus}},\ }\href@noop {} {\bibfield  {journal} {\bibinfo
  {journal} {{JETP} Letters}\ }\textbf {\bibinfo {volume} {60}},\ \bibinfo
  {pages} {206} (\bibinfo {year} {1994})}\BibitemShut {NoStop}%
\bibitem [{\citenamefont {Pikus}\ and\ \citenamefont
  {Pikus}(1995)}]{pikus:1995}%
  \BibitemOpen
  \bibfield  {author} {\bibinfo {author} {\bibfnamefont {F.~G.}\ \bibnamefont
  {Pikus}}\ and\ \bibinfo {author} {\bibfnamefont {G.~E.}\ \bibnamefont
  {Pikus}},\ }\href@noop {} {\bibfield  {journal} {\bibinfo  {journal}
  {Physical Review B}\ }\textbf {\bibinfo {volume} {51}},\ \bibinfo {pages}
  {16928} (\bibinfo {year} {1995})}\BibitemShut {NoStop}%
\bibitem [{\citenamefont {Luo}\ \emph {et~al.}(1990)\citenamefont {Luo},
  \citenamefont {Munekata}, \citenamefont {Fang},\ and\ \citenamefont
  {Stiles}}]{luo:1990}%
  \BibitemOpen
  \bibfield  {author} {\bibinfo {author} {\bibfnamefont {J.}~\bibnamefont
  {Luo}}, \bibinfo {author} {\bibfnamefont {H.}~\bibnamefont {Munekata}},
  \bibinfo {author} {\bibfnamefont {F.~F.}\ \bibnamefont {Fang}}, \ and\
  \bibinfo {author} {\bibfnamefont {P.~J.}\ \bibnamefont {Stiles}},\
  }\href@noop {} {\bibfield  {journal} {\bibinfo  {journal} {Physical Review
  B}\ }\textbf {\bibinfo {volume} {41}},\ \bibinfo {pages} {7685} (\bibinfo
  {year} {1990})}\BibitemShut {NoStop}%
\bibitem [{\citenamefont {Papadakis}\ \emph {et~al.}(1999)\citenamefont
  {Papadakis}, \citenamefont {De~Poortere}, \citenamefont {Manoharan},
  \citenamefont {Shayegan},\ and\ \citenamefont {Winkler}}]{papadakis:1999}%
  \BibitemOpen
  \bibfield  {author} {\bibinfo {author} {\bibfnamefont {S.~J.}\ \bibnamefont
  {Papadakis}}, \bibinfo {author} {\bibfnamefont {E.~P.}\ \bibnamefont
  {De~Poortere}}, \bibinfo {author} {\bibfnamefont {H.~C.}\ \bibnamefont
  {Manoharan}}, \bibinfo {author} {\bibfnamefont {M.}~\bibnamefont {Shayegan}},
  \ and\ \bibinfo {author} {\bibfnamefont {R.}~\bibnamefont {Winkler}},\
  }\href@noop {} {\bibfield  {journal} {\bibinfo  {journal} {Science}\ }\textbf
  {\bibinfo {volume} {283}},\ \bibinfo {pages} {2056} (\bibinfo {year}
  {1999})}\BibitemShut {NoStop}%
\bibitem [{\citenamefont {Grundler}(2000)}]{grundler:2000}%
  \BibitemOpen
  \bibfield  {author} {\bibinfo {author} {\bibfnamefont {D.}~\bibnamefont
  {Grundler}},\ }\href@noop {} {\bibfield  {journal} {\bibinfo  {journal}
  {Physical Review Letters}\ }\textbf {\bibinfo {volume} {84}},\ \bibinfo
  {pages} {6074} (\bibinfo {year} {2000})}\BibitemShut {NoStop}%
\bibitem [{\citenamefont {Bergmann}(1984)}]{bergmannreview}%
  \BibitemOpen
  \bibfield  {author} {\bibinfo {author} {\bibfnamefont {G.}~\bibnamefont
  {Bergmann}},\ }\href@noop {} {\bibfield  {journal} {\bibinfo  {journal}
  {Physics Reports}\ }\textbf {\bibinfo {volume} {107}},\ \bibinfo {pages} {1}
  (\bibinfo {year} {1984})}\BibitemShut {NoStop}%
\bibitem [{\citenamefont {Altschuler}\ and\ \citenamefont
  {Aronov}(1985)}]{AAreview}%
  \BibitemOpen
  \bibfield  {author} {\bibinfo {author} {\bibfnamefont {B.~L.}\ \bibnamefont
  {Altschuler}}\ and\ \bibinfo {author} {\bibfnamefont {A.~G.}\ \bibnamefont
  {Aronov}},\ }\href@noop {} {\bibfield  {journal} {\bibinfo  {journal}
  {Electron-Electron Interactions in Disordered Systems, Edited by A. L. Efros
  and M. Pollak}\ } (\bibinfo {year} {1985})}\BibitemShut {NoStop}%
\bibitem [{\citenamefont {Knap}\ \emph {et~al.}(1996)\citenamefont {Knap},
  \citenamefont {Skierbiszewski}, \citenamefont {Zduniak}, \citenamefont
  {Litwin-Staszewska}, \citenamefont {Bertho}, \citenamefont {Kobbi},
  \citenamefont {Robert}, \citenamefont {Pikus}, \citenamefont {Pikus},
  \citenamefont {Iordanskii}, \citenamefont {Mosser}, \citenamefont
  {Zekentes},\ and\ \citenamefont {Lyanda-Geller}}]{knap:1996}%
  \BibitemOpen
  \bibfield  {author} {\bibinfo {author} {\bibfnamefont {W.}~\bibnamefont
  {Knap}}, \bibinfo {author} {\bibfnamefont {C.}~\bibnamefont
  {Skierbiszewski}}, \bibinfo {author} {\bibfnamefont {A.}~\bibnamefont
  {Zduniak}}, \bibinfo {author} {\bibfnamefont {E.}~\bibnamefont
  {Litwin-Staszewska}}, \bibinfo {author} {\bibfnamefont {D.}~\bibnamefont
  {Bertho}}, \bibinfo {author} {\bibfnamefont {F.}~\bibnamefont {Kobbi}},
  \bibinfo {author} {\bibfnamefont {J.~L.}\ \bibnamefont {Robert}}, \bibinfo
  {author} {\bibfnamefont {G.~E.}\ \bibnamefont {Pikus}}, \bibinfo {author}
  {\bibfnamefont {F.~G.}\ \bibnamefont {Pikus}}, \bibinfo {author}
  {\bibfnamefont {S.~V.}\ \bibnamefont {Iordanskii}}, \bibinfo {author}
  {\bibfnamefont {V.}~\bibnamefont {Mosser}}, \bibinfo {author} {\bibfnamefont
  {K.}~\bibnamefont {Zekentes}}, \ and\ \bibinfo {author} {\bibfnamefont
  {Y.~B.}\ \bibnamefont {Lyanda-Geller}},\ }\href@noop {} {\bibfield  {journal}
  {\bibinfo  {journal} {Physical Review B}\ }\textbf {\bibinfo {volume} {53}},\
  \bibinfo {pages} {3912} (\bibinfo {year} {1996})}\BibitemShut {NoStop}%
\bibitem [{\citenamefont {Miller}\ \emph {et~al.}(2003)\citenamefont {Miller},
  \citenamefont {Zumb\"hl}, \citenamefont {Marcus}, \citenamefont
  {Lyanda-Geller}, \citenamefont {Goldhaber-Gordon}, \citenamefont {Campman},\
  and\ \citenamefont {Gossard}}]{miller:2003}%
  \BibitemOpen
  \bibfield  {author} {\bibinfo {author} {\bibfnamefont {J.~B.}\ \bibnamefont
  {Miller}}, \bibinfo {author} {\bibfnamefont {D.~M.}\ \bibnamefont
  {Zumb\"hl}}, \bibinfo {author} {\bibfnamefont {C.~M.}\ \bibnamefont
  {Marcus}}, \bibinfo {author} {\bibfnamefont {Y.~B.}\ \bibnamefont
  {Lyanda-Geller}}, \bibinfo {author} {\bibfnamefont {D.}~\bibnamefont
  {Goldhaber-Gordon}}, \bibinfo {author} {\bibfnamefont {K.}~\bibnamefont
  {Campman}}, \ and\ \bibinfo {author} {\bibfnamefont {A.~C.}\ \bibnamefont
  {Gossard}},\ }\href {\doibase 10.1103/PhysRevLett.90.076807} {\bibfield
  {journal} {\bibinfo  {journal} {Physical Review Letters}\ }\textbf {\bibinfo
  {volume} {90}},\ \bibinfo {pages} {076807} (\bibinfo {year}
  {2003})}\BibitemShut {NoStop}%
\bibitem [{\citenamefont {Calsaverini}\ \emph {et~al.}(2008)\citenamefont
  {Calsaverini}, \citenamefont {Bernardes}, \citenamefont {Egues},\ and\
  \citenamefont {Loss}}]{calsaverini:2008}%
  \BibitemOpen
  \bibfield  {author} {\bibinfo {author} {\bibfnamefont {R.~S.}\ \bibnamefont
  {Calsaverini}}, \bibinfo {author} {\bibfnamefont {E.}~\bibnamefont
  {Bernardes}}, \bibinfo {author} {\bibfnamefont {J.~C.}\ \bibnamefont
  {Egues}}, \ and\ \bibinfo {author} {\bibfnamefont {D.}~\bibnamefont {Loss}},\
  }\href@noop {} {\bibfield  {journal} {\bibinfo  {journal} {Physical Review
  B}\ }\textbf {\bibinfo {volume} {78}},\ \bibinfo {pages} {155313} (\bibinfo
  {year} {2008})}\BibitemShut {NoStop}%
\bibitem [{\citenamefont {Krich}\ and\ \citenamefont
  {Halperin}(2007)}]{krich:2007}%
  \BibitemOpen
  \bibfield  {author} {\bibinfo {author} {\bibfnamefont {J.}~\bibnamefont
  {Krich}}\ and\ \bibinfo {author} {\bibfnamefont {B.}~\bibnamefont
  {Halperin}},\ }\href {\doibase 10.1103/PhysRevLett.98.226802} {\bibfield
  {journal} {\bibinfo  {journal} {Physical Review Letters}\ }\textbf {\bibinfo
  {volume} {98}},\ \bibinfo {pages} {226802} (\bibinfo {year}
  {2007})}\BibitemShut {NoStop}%
\bibitem [{\citenamefont {Walser}\ \emph
  {et~al.}(2012{\natexlab{b}})\citenamefont {Walser}, \citenamefont
  {Siegenthaler}, \citenamefont {Lechner}, \citenamefont {Schuh}, \citenamefont
  {Ganichev}, \citenamefont {Wegscheider},\ and\ \citenamefont
  {Salis}}]{walser:2012_2}%
  \BibitemOpen
  \bibfield  {author} {\bibinfo {author} {\bibfnamefont {M.~P.}\ \bibnamefont
  {Walser}}, \bibinfo {author} {\bibfnamefont {U.}~\bibnamefont
  {Siegenthaler}}, \bibinfo {author} {\bibfnamefont {V.}~\bibnamefont
  {Lechner}}, \bibinfo {author} {\bibfnamefont {D.}~\bibnamefont {Schuh}},
  \bibinfo {author} {\bibfnamefont {S.~D.}\ \bibnamefont {Ganichev}}, \bibinfo
  {author} {\bibfnamefont {W.}~\bibnamefont {Wegscheider}}, \ and\ \bibinfo
  {author} {\bibfnamefont {G.}~\bibnamefont {Salis}},\ }\href {\doibase
  10.1103/PhysRevB.86.195309} {\bibfield  {journal} {\bibinfo  {journal}
  {Physical Review B}\ }\textbf {\bibinfo {volume} {86}},\ \bibinfo {pages}
  {195309} (\bibinfo {year} {2012}{\natexlab{b}})}\BibitemShut {NoStop}%
\bibitem [{\citenamefont {Fu}\ \emph {et~al.}(2016)\citenamefont {Fu},
  \citenamefont {Penteado}, \citenamefont {Hachiya}, \citenamefont {Loss},\
  and\ \citenamefont {Egues}}]{fu:2015}%
  \BibitemOpen
  \bibfield  {author} {\bibinfo {author} {\bibfnamefont {J.}~\bibnamefont
  {Fu}}, \bibinfo {author} {\bibfnamefont {P.~H.}\ \bibnamefont {Penteado}},
  \bibinfo {author} {\bibfnamefont {M.~O.}\ \bibnamefont {Hachiya}}, \bibinfo
  {author} {\bibfnamefont {D.}~\bibnamefont {Loss}}, \ and\ \bibinfo {author}
  {\bibfnamefont {J.~C.}\ \bibnamefont {Egues}},\ }\href {\doibase
  10.1103/PhysRevLett.117.226401} {\bibfield  {journal} {\bibinfo  {journal}
  {Phys. Rev. Lett.}\ }\textbf {\bibinfo {volume} {117}},\ \bibinfo {pages}
  {226401} (\bibinfo {year} {2016})}\BibitemShut {NoStop}%
\end{thebibliography}
\end{document}